\documentclass[11pt,a4paper]{article} 
\usepackage[numbers,sort&compress]{natbib}
\usepackage{amsmath}
\usepackage{amssymb}
\usepackage{authblk}
\usepackage{graphicx}
\usepackage{braket}
\usepackage{caption}
\usepackage{braket}
\title{Field test of entanglement swapping over 100-km optical fiber with independent 1-GHz-clock sequential time-bin entangled photon-pair  sources}

\author[1,2,3,*]{Qi-Chao Sun}
\author[1,2,*]{Yang-Fan Jiang}
\author[1,2,*]{Ya-Li Mao}
\author[4]{Li-Xing You}
\author[5]{Wei Zhang}
\author[4]{Wei-Jun Zhang}
\author[1,2]{Xiao Jiang}
\author[1,2]{Teng-Yun Chen}
\author[4]{Hao Li}
\author[5]{Yi-Dong Huang}
\author[3]{Xian-Feng Chen}
\author[4]{Zhen Wang}
\author[1,2]{Jingyun Fan}
\author[1,2]{Qiang Zhang}
\author[1,2]{Jian-Wei Pan}

\affil[1]{National Laboratory for Physical Sciences at Microscale and Department of Modern Physics, Shanghai Branch, University of Science and Technology of China, Hefei, Anhui 230026, China}
\affil[2]{CAS Center for Excellence and Synergetic Innovation Center in Quantum Information and Quantum Physics, Shanghai Branch, University of Science and Technology of China, Hefei, Anhui 230026, China}
\affil[3]{School of Physics and Astronomy, Shanghai Jiao Tong University, Shanghai, 200240, China}
\affil[4]{State Key Laboratory of Functional Materials for Informatics, Shanghai Institute of Microsystem and Information Technology, Chinese Academy of Sciences, Shanghai 200050, China}
\affil[5]{Tsinghua National Laboratory for Information Science and Technology, Department of Electronic Engineering, Tsinghua University, Beijing 100084, China}
\affil[*]{These authors contributed equally to this work}

\begin{document}

\maketitle

\begin{abstract}
Realizing long distance entanglement swapping with independent sources in the real-world condition is important for both future quantum network and fundamental study of quantum theory. Currently, demonstration over a few of tens kilometer underground optical fiber has been achieved. However, future applications demand entanglement swapping over longer distance with more complicated environment. We exploit two independent 1-GHz-clock sequential time-bin entangled photon-pair sources, develop several automatic stability controls, and successfully implement a field test of entanglement swapping over more than 100-km optical fiber link including coiled, underground and suspended optical fibers. Our result verifies the feasibility of such technologies for long distance quantum network and for many interesting quantum information experiments. 
\end{abstract}

Entanglement swapping~\cite{ZukowskiEntanglementSwapping,Pan1998EntanglementSwap} is a unique feature of quantum physics. By entangling two independent parties that have never interacted before, entanglement swapping has been used in the study of physics foundations such as  nonlocality~\cite{ZukowskiEntanglementSwapping,Branciard2010Bilocal} and wave-particle duality~\cite{Peres2000EntSwapDelCho}. It is also a central element in quantum newtwork~\cite{Cirac:QuantumNetwork,Kimble2008QuanInt}, appearing in the form of quantum relay~\cite{Waks2002QuanRel,Jacobs2002QuanRel,Collins2003QuanRel} and quantum repeater~\cite{Briegel1998QuantumRepeater,DLCZ}. The integrity of an experimental realization of entanglemnet swapping is ensured only by satisfying these criteria: proper causal disconnection between relevant events ~\cite{Ma2012SwappingDelayChoice, Hensen2015LoopholeFreeBT}, and independent quantum sources without common past~\cite{Branciard2010Bilocal}.

\begin{figure}[!htbp]
\centering
\includegraphics[width=\linewidth]{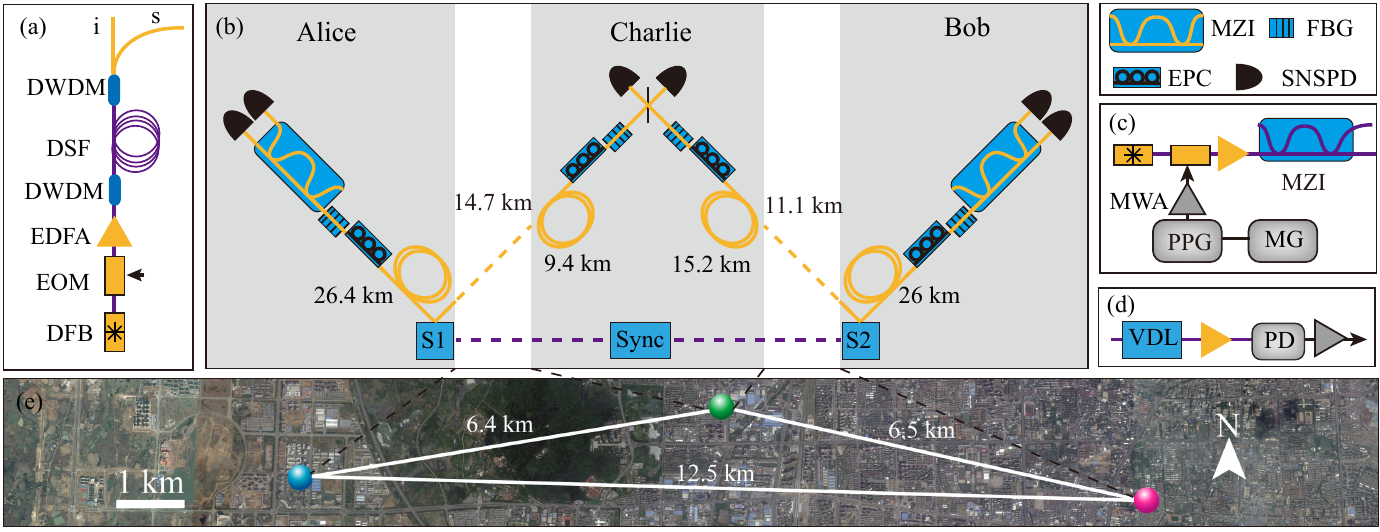}
\caption{Scheme of the entanglement swapping experiment. (a) Setup of sequential time-bin entangled photon-pair source. DFB, distributed feedback laser; EOM, electro-optic modulator; EDFA, erbium doped fiber amplifier; DWDM, dense wavelength division multiplexing filter; DSF, dispersion shifted fiber. (b) Experimental realization. Two sequential time-bin entangled photon-pair sources, S1 and S2,  are placed in nodes Alice and Bob, respectively. Each of them keeps the idler photons in coiled fibers and distributed the signal photons to Charlie through deployed optical fiber. The yellow dash lines represent deployed fibers and the circles represent the coiled fibers. The total transmission loss of  the 77-km coiled optical fiber is about 16~dB, while that of the deployed optical fiber in Alice-Charlie and Bob-Charlie links are 6~dB and 7~dB, respectively. To synchronize the two sources, Charlie prepares 1-GHz laser pulses (Sync) and sends them to Alice and Bob through deployed optical fiber (represented by the purple dash lines) to generate the driven signal for their EOMs. The setup  MZI, unbalanced Mach-Zehnder interferometer; EPC, electronic-controlled polarization controller; FBG, fiber Bragg grating; SNSPD, superconducting nanowire single photon detector. (c) Setup of synchronization signal generation. Charlie first generates laser pulses with  repetition rate of  500 MHz, and then doubles them using an MZI with 1-ns path difference. MG, microwave generator; PPG, pulse pattern generator; MWA, microwave amplifier. (d) Setup to generate the driven signal for EOM in (a). VDL, variable delay line; PD, photodiode. (e), satellite image of the experimental nodes.}
\label{fig:scheme}
\end{figure}

Driven by the application of future quantum network, there has been a significant progress in experimental entanglement swapping since its first experimental demonstration \cite{Pan1998EntanglementSwap}. Quantun interference with independent sources was addressed in a number of experimental settings ~\cite{Yang2006IndSource,Kaltenbaek2006IndSource,Halder2007SwappingCWPump,Kaltenbaek2009HFEntSwap,Stevenson2013TeleLED}.Quantum relay was simulated with coiled optical fibers in a laboratory environment ~\cite{Riedmatten2004TeleportationQuantumRelay}. Recently, entanglement swapping and quantum teleportation were realized in both free space and optical fiber link with about 100-km distance~\cite{Yin2015Teleportation100km,Ma2012Teleportation143km,Herbst2015EntSwap143, Hiroki2015Teleportation}, in which  the quantum sources shared a common past, the Bell state measurement (BSM) were performed locally, and the swapped (teleported) photonic qubits were sent afterwards over long distance for analysis. More recently,  several teams succeeded in entanglement swapping and teleportation with high integrity over a fiber network (about $10$ km) in the real world, in which  they  overcomed the challenge in removing the distinguishability between photons from separate quantum sources by defeating the noise in the real world~\cite{Hensen2015LoopholeFreeBT, Sun2016Tel, Raju2016Tele, Sun2016ES}. To date, there is no report on entanglement swapping or quantum teleportation with suspended optical fiber, which is more susceptible to environment but unavoidable for applications in optical fiber network.

Here, we present an implementation of entanglement swapping in an inter-city quantum network, which is composed of about 77-km optical fiber inside the lab, 25-km optical fiber outside of the lab but kept underground, and 1-km optical fiber suspended in the air outside of the lab to account for various types of noise mechanisms in the real world.

The schematic of the sequential time-bin entangled photon-pair source~\cite{Riedmatten2004Sequential,Zhang2008Seq,Hiroki2009Swapping} is depicted in Fig.1(a). We carve the continuous wave output of a distributed feedback laser (with coherence time $\tau_c  = 300 \mu s$) periodically into pulses at a rate of 1 GHz with an electro-optical modulator (EOM). The sequential laser pulses differ with $\tau$ = 1 ns and a phase difference of $\theta = 2\pi\nu\tau$. After amplification with an erbium doped fiber amplifier and spectral filtering with dense wavelength division multiplexing filters (DWDM), the laser pulses are fed into  a 300-m dispersion shifted fiber immersed  the liquid nitrogen to produce photon-pairs via spontaneous four-wave mixing. The quantum state of a produced photon-pair is given by $\ket{\Theta}=\frac{1}{\sqrt{n}}\sum_{k=0}^{n-1}e^{2ik\theta}\ket{t_k}_s\ket{t_k}_i$, where $t_k=k\tau$ represents time bin k, and  $s$ and $i$ are for idler (1,555.73~nm)/signal  (1,549.36~nm) photons. We single out  the signal/idler photons with cascaded DWDM filters with pump photons suppressed by 115~dB.

\begin{figure}[!htbp]
\centering
\includegraphics[width=\linewidth]{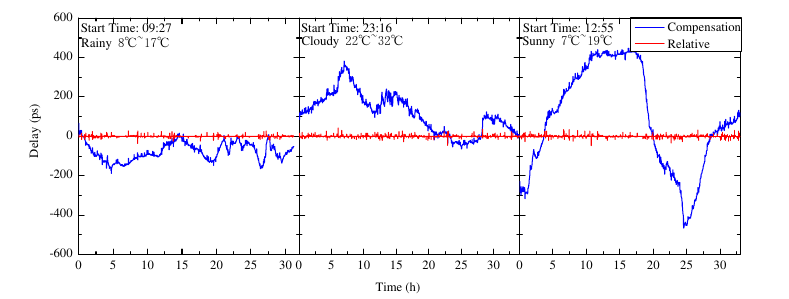}
\caption{Typical delay compensation (blue line) and relative delay between arrival time of photons from Alice and Bob  (red line) under different weather conditions. Measured by a TDC with time resolution of 4~ps,  the standard deviations of the relative delay in (a)-(d) are 6.1~ps, 6.5~ps, 6.0~ps, and 6.7~ps, respectively.  }
\label{fig:delay}
\end{figure}

The schematic of the entanglement swapping experiment  is shown in Fig.\ref{fig:scheme} (b), with(out) gray shaded areas indicating indoor (outdoor) environment. The two quantum sources are placed 12.5 km apart at nodes Alice and Bob, respectively,  and BSM is performed at the third node (Charlie) between them, as shown in Fig.\ref{fig:scheme} (e).  Alice and Bob hold idler photons with 26-km coiled fibers while sending signal photons to Charlie. We specifically keep 1-km optical fiber cable between Bob and Charlie  suspended in the air and exposed to sunlight and wind to face inclement weather conditions. The rest deployed fibers are kept underground.  The total length of the optical fiber is 103 km with a total  transmission loss of $\sim29\ \mathrm{dB}$.  Upon detection,  both idler and signal photons are passed through 4-GHz fiber Bragg gratings, which is about half of the bandwidth of the pump pulses and is therefore sufficient to eliminate the frequency correlation between the twin photons\citep{Sun2016Tel}. To synchronize the operation of each node in the quantum network, Charlie keeps a master clock which sends pulses to Alice and Bob at 1 GHz, which is used to drive the EOM to synchronize independent quantum sources (see Fig.\ref{fig:scheme} (c,d)).

In the BSM, it is critical that the signal photons sent by Alice and Bob arrive at the 50:50 beam splitter (BS) simultaneously. However, the arrival time changes drastically due to the fluctuation of effective length of the optical fiber link which is subjected to the influence in the real world. As shown in Fig.~\ref{fig:delay}, the typical peak-to-peak delays between arrival times of photons from Alice and Bob changes are 200~ps, 500~ps, and 1000~ps in rainy days, cloudy days, and sunny days, respectively, which are much larger than the coherent time of signal photons ($\sim110\ \mathrm{ps}$). We use the difference between the arrival times of signal photons from Alice and Bob as error signals and feed them into a delay line to suppress  the relative delay to 6~ps under all weather conditions, which is $\ll \sim110\ \mathrm{ps}$ to ensure high interference visibility.  In addition,  for each channel, an electronic-controlled polarization controller is used to compensate the polarization fluctuation caused by optical fiber.

\begin{figure}[!htbp]
\centering
\includegraphics[width=8 cm]{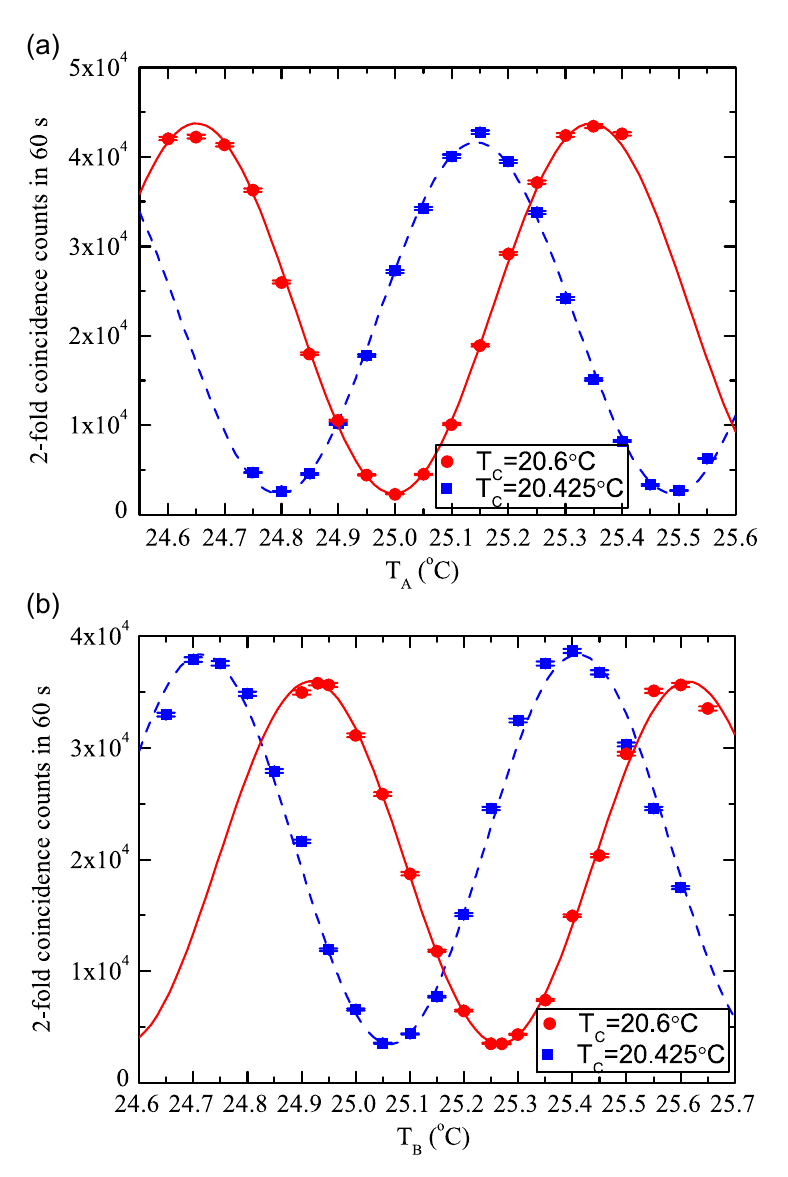}
\caption{Two-fold coincidence counts of the sequential time-bin entangled photon-pairs distributed by Alice (a) and Bob (b) as functions of temperature of the MZIs. The measurement results are represented by squares and circles corresponding to Charlie's MZI at  temperature of $20.425\ ^\circ C$ and $20.6\ ^\circ C$, respectively. The error bars indicate one standard deviation calculated from measured counts assuming Poissonian detection statistics. The visibility of the fitted sinusoidal curves for the squares (circles) is  $(89.8\pm0.6)\%$ ( $(89.5\pm1.2)\%$) and  $(83.5\pm1.5)\%$ ($(82.3\pm0.8)\%$) for (a) and (b), respectively. }
\label{fig:source}
\end{figure}

Before  performing entanglement swapping, we characterize the quantum source of Alice (Bob) after entanglement distribution using  the Franson-type interferometer. At nodes Alice (Bob) and Charlie, the idler and signal photon are fed into an unbalanced Mach-Zehnder interferometer (MZI) with an arm difference of 1-ns.  The interference term in one output of the MZI can be written as $\sum_{k=1}^{n-1}e^{2ik\theta}(1+e^{i(\theta_s+\theta_i-2\theta)})\ket{t_k}_s\ket{t_k}_i$, where $\theta_s$ and $\theta_i$ are relative phases induced by MZIs for single and idler photons, respectively. The outputs of MZIs are detected by superconducting nanowire single photon detectors (SNSPD) and the detection results are recorded by time-to-digital converters (TDC) with a 4-ps resolution and analyzed in real-time. The TDCs are synchronized with 10 MHz clocks at Charlie's node. Fig.\ref{fig:source} shows that the measured two-fold coincidence counts between Alice (Bob) and Charlie change sinusoidally as a function of phase (which is dialed by sweeping the temperature of MZI), with a visibility of  $(89.8\pm0.5)\%$ for Alice's source and $(82.9\pm1.2)\%$ for Bob's source, respectively. 

In the entanglement swapping experiment, Charlie performs BSM with signal photons sent by Alice and Bob by interfering them on a  BS. By detecting them with two SNSPDs and within a delay of 1 ns (for the same time-bin or adjacent time-bins), the  overall state of the two entangled photon-pairs, $\ket{\Gamma}_{1,2}=\frac{1}{n}(\sum_{k=0}^{n-1}e^{2ik\theta}\ket{t_k}_{1s}\ket{t_k}_{1i})\otimes(\sum_{l=0}^{n-1}e^{2il\theta}\ket{t_l}_{2s}\ket{t_l}_{2i})$, can be cast into
\begin{equation*}
\begin{split}
\ket{\Gamma}_{1,2}\rightarrow&\frac{1}{n}\{\sum_{k=0}^{n-1}\frac{e^{4ik\theta}}{\sqrt{2}n}\ket{t_k}_{1i}\ket{t_k}_{2i}(\ket{\Phi^+}_{s,k}+\ket{\Phi^-}_{s,k})+\\
&\sum_{k=0}^{n-2}e^{i(4k+2)\theta}[\frac{1}{\sqrt{2}}(\ket{t_k}_{1i}\ket{t_{k+1}}_{2i}+\ket{t_{k+1}}_{1i}\ket{t_k}_{2i})\ket{\Psi^+}_{s,k}\\
&\frac{1}{\sqrt{2}}(\ket{t_k}_{1i}\ket{t_{k+1}}_{2i}-\ket{t_{k+1}}_{1i}\ket{t_k}_{2i})\ket{\Psi^-}_{s,k}]\},
\end{split}
\end{equation*}
where  the four Bell state sets are given by   $\{\ket{\Psi^\pm}_{s,k}=\frac{1}{\sqrt{2}}(\ket{t_k}_{1s}\ket{t_{k+1}}_{2s}\pm\ket{t_{k+1}}_{1s}\ket{t_k}_{2s})\}$ and $\{ \ket{\Phi^\pm}_{s,k}=\frac{1}{\sqrt{2}}(\ket{t_k}_{1s}\ket{t_k}_{2s}\pm\ket{t_{k+1}}_{1s}\ket{t_{k+1}}_{2s})\}$.

\begin{figure}[!htbp]
\centering
\includegraphics[width=8 cm]{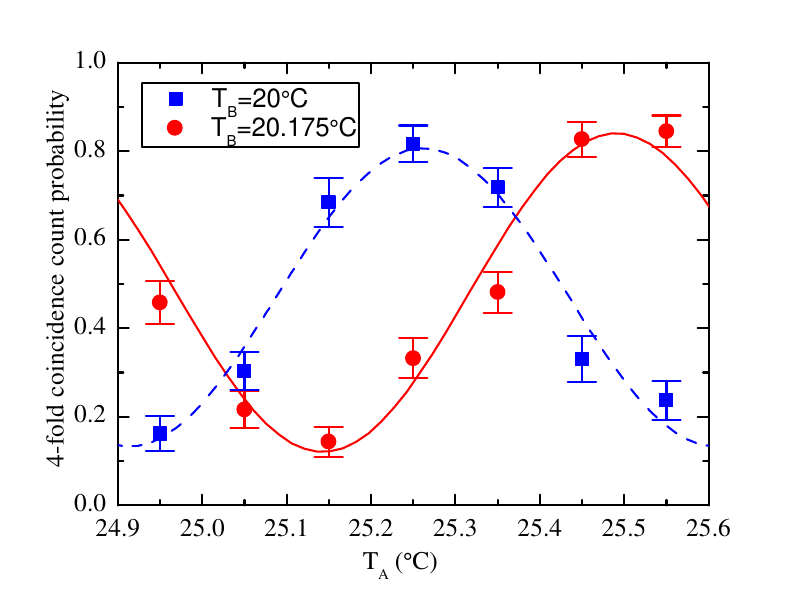}
\caption{Four-fold coincidence count probabilities as a function of the temperature of Alice's MZI. The error bars indicate one standard deviation calculated from measured counts assuming Poissonian detection statistics. Each data point is accumulated for more than 30 h. The visibility of the fitted curve is  $(74.8\pm8.7)\%$ and  $(71.7\pm7.2)\%$ for measured results with $T_B=20.175\ ^\circ C$  and $T_B=20\ ^\circ C$, respectively.}
\label{fig:IntCur}
\end{figure}

 The Bell states $\ket{\Phi^\pm}_{s,k}$ correspond to cases that the two signal photons are output from the same port of the BS and in the same time bin, which can not be discriminated using linear optics. The Bell states $\ket{\Psi^+}_{s,k}$ and $\ket{\Psi^-}_{s,k}$ correspond to cases that the two signal photons are output in time bins with 1-ns delay, while from the same port and different ports of the BS, respectively. In our experiment, the recovering time of the SNSPD is about 40~ns, much longer than the time delay of two consecutive time bins. So only the Bell states $\{\ket{\Psi_k^-}_{s,k}\}$ are discriminated in BSM. As a result, their twin photons are projected to entangled quantum state $\ket{\Psi^-}_{i,k}=\frac{1}{\sqrt{2}}(\ket{t_k}_{1i}\ket{t_{k+1}}_{2i}-\ket{t_{k+1}}_{1i}\ket{t_k}_{2i})$. To verify the entanglement swapping,  both Alice and Bob use a MZI with  1-ns path difference followed by SNSPDs and implement a conditioned Franson-type measurment for time-bin entangled state. The experimental result is shown in  Fig.~\ref{fig:IntCur}. The four-fold coincidence counts show a clear interference fringe and the average visibility of the fitted curves is $(73.2\pm5.6)\%$.  If we assume that the two photons are in a Werner state, we can show that the lower bound of visibility to demonstrate entanglement is 1/3~\cite{Peres1996Separability}. The visibility achieved in our experiment clearly exceeds this bound. 

In our experiment, the total transmission loss of the optical fiber link is about 10 to 20~dB higher than previous field tests with independent sources. By setting the interval between adjacent coherent pump laser pulses equal to that of conventional time-bin entangled photon-pair source, the event rate of experiment with sequential time-bin entangled photon pair source can be increased by 3 times. We achieve a four-fold count rate in energy basis of $\sim3/\mathrm{h}$ with 1-GHz pump laser pulses.  The relative low visibility of the entanglement created in entanglement swapping is mainly attributed to the imperfect sequential time-bin entangled photon-pair sources, which upper bounds the visibility to $\sim74\%$ ( product of the visibility of two entanglement sources). In our experiment, the average photon-pair number per time bin is $\mu\approx0.023$ for both sources, which can decrease the visibility to  $\sim96\%$ ($V_d\approx1/(1+2\mu)$). The wavelength of the CW laser is controlled with stability of $0.18\ \mathrm{pm}$, resulting in a fluctuation of the relative phase in the entangled state. Assuming the temperature fluctuation follows a Gaussian distribution, it can decrease the visibility to  $\sim96\%$. The  distortion of driven signals can also decrease the visibility. The main reason of distortion is the limited bandwidth of PD, which is 45 GHz and 10 GHz for Alice and Bob, respectively. So by employing a 45-GHz PD in Bob's source, it is possible that the visibility of the swapped entanglement can be increased to $\sim80\%$ so that it can be used to demonstrate quantum key distribution~\cite{Sun2016ES}.

In summary,  we have demonstrated entanglement swapping with two independent sources 12.5 km apart using 103-km optical fiber.  Compared with previous experiments with independent sources, we have increased the length of optical fiber from metropolitan distance to inter-city distance. The transmission loss and stability of the optical fiber channel in our experiment is enough to match those of more than 100-km typical underground deployed optical fiber~\cite{LossDepolyedFiber}.  So, our results show that realizing entanglement swapping between two city is technically feasible. Moreover, the configuration of our experiment allows the space-like separation between any two measurements of those performed in the three nodes, and various of time-space relation can be achieved by combining both coiled optical fiber and deployed optical fiber.  This distinguish feature together with the independent sources make our setup a promising platform for many interesting fundamental tests.

\section*{Funding Information}
National Fundamental Research Program (under Grants No. 2013CB336800); National Natural Science Foundation of China; Chinese Academy of Science; 10000-Plan of Shandong Province; Quantum Ctek Co., Ltd. 

\section*{Acknowledgments}
The authors thank M.-H Li, B. Wang, X.-L. Wang, and Y. Liu for enlightening discussions.

\end{document}